\documentclass[twocolumn,prd,aps,showpacs,preprintnumbers,nofootinbib]{revtex4}

\usepackage{amsmath,amsfonts,amssymb}
\usepackage{wrapfig}           
\usepackage{graphicx}          
\usepackage{color}             
\usepackage{dcolumn}           
\usepackage{psfrag}
\usepackage{pstricks,pst-plot} 

\usepackage{t1enc}
\usepackage[english]{babel}   
\usepackage[latin1]{inputenc}

\allowdisplaybreaks

\newcommand{\be}{\begin{equation}}
\newcommand{\ee}{\end{equation}}
\newcommand{\bea}{\begin{eqnarray}}
\newcommand{\eea}{\end{eqnarray}}

\newcommand{\ben}{\begin{eqnarray*}}
\newcommand{\een}{\end{eqnarray*}}

\newcommand{\bs}{\begin{subequations}}
\newcommand{\es}{\end{subequations}}

\newcommand{\vek}[1]{\boldsymbol{#1}}

\begin{document}

\author{Achamveedu Gopakumar}
\email{A.Gopakumar@uni-jena.de}
\author{Christian K\"onigsd\"orffer}
\email{C.Koenigsdoerffer@uni-jena.de}

\affiliation{
Theoretisch-Physikalisches Institut,
Friedrich-Schiller-Universit\"at, 
Max-Wien-Platz 1, 
07743 Jena, 
Germany}

\title{The deterministic nature of conservative post-Newtonian
accurate dynamics of compact binaries with
leading order spin-orbit interaction}

\date{\today}
     
\begin{abstract}
We formally show that the conservative 
second post-Newtonian (PN) accurate
dynamics of spinning compact binaries moving in eccentric orbits,
when spin effects are restricted to the leading order
spin-orbit interaction cannot be chaotic for the 
following two distinct cases: 
(i) the binary consists of compact objects of arbitrary mass, 
where only one of them is spinning with an arbitrary spin and 
(ii) the binary consists of equal mass compact objects, 
having two arbitrary spins.
We rest our arguments on the recent determination of 
PN accurate Keplerian-type parametric solutions 
to the above cases, indicating that the PN accurate dynamics
is integrable in these two situations. 
We compare predictions of our case (i) 
with those from a numerical investigation
of an equivalent scenario that observed
chaos in the associated dynamics.
We also present possible reasons for the discrepancies.
\end{abstract}


\maketitle

\section{Introduction}

Inspiralling binaries consisting of spinning compact objects
of arbitrary mass ratio are astrophysically interesting systems.
These binaries are the most promising targets for the 
first direct detection of gravitational radiation
by ground-based 
laser interferometric gravitational-wave detectors \cite{VK04}.
The radio-wave detections of spinning compact binaries 
are heavily sought after as they provide unique 
laboratories for relativistic gravity and plasma physics \cite{L04}.
The orbital dynamics of such compact binaries are 
well described by the post-Newtonian (PN)
approximation to general relativity.
The PN approximation to general relativity
allows one to express the equations of motion for a compact binary
as corrections to Newtonian equations of motion
in powers of $(v/c)^2 \sim GM/(c^2 R)$, where $v$, $M$, and $R$ are
the characteristic orbital velocity, 
the total mass and the typical orbital separation of the binary.
To extract lots of astrophysical informations from these binaries,
both using gravitational and radio wave observations,
it is understood that the compact binary dynamics
should be known at least to 
second post-Newtonian (2PN) order \cite{PW95,MBSP04}.
The 2PN order gives $(v/c)^4$ corrections
to Newtonian equations of motion and the orbital dynamics is still
conservative at this order.
In the case of spinning compact binaries,
the dynamics is determined,
not only by the orbital equations of motion for these objects,
but also by the precessional equations for the orbital plane and
the spin vectors themselves \cite{Schaefer_grav_effects_2004}.
The dynamics of these systems are required to be 
deterministic, at least during the observational window,
for making any meaningful astrophysical measurements.

Few years ago, it was argued that spin-orbit and spin-spin contributions 
to the conservative PN dynamics will force the orbits 
of compact binaries to become chaotic \cite{JL00}.
The ``chaos'' in the conservative dynamics of 
spinning compact binaries were identified with the aid of phase-space
trajectories that form fractal basin boundaries.
These trajectories were computed by numerically integrating 
2PN accurate equations, describing the compact binary dynamics
in harmonic gauge, available in Ref.~\cite{K95}.
It was reasoned that compact binaries, having initial conditions near
the fractal basin boundaries will have unpredictable evolution.
Therefore, it was argued that spinning compact binaries,
evolving under PN accurate orbital dynamics, will have 
unpredictable gravitational waveforms during the inspiral.
This will make the deployment of ``matched filtering'' and
hence the detection of these binaries, via gravitational 
radiation, an impossible task.
The above claim was questioned in Ref.~\cite{SR01}.
These authors computed the divergence of nearby 
phase space trajectories of spinning compact binaries,
with the aid of Lyapunov exponents,
evolving under the same PN dynamics, as investigated in Ref.~\cite{JL00}
and found no evidence for any chaotic behavior.
However, the above analysis was strongly criticized
in Ref.~\cite{CL03_prd}, based on arguments that in Ref.~\cite{SR01}
the ``proper'' Lyapunov exponents were not computed.
Further, the subtleties involved in the use of Lyapunov exponents
in general relativity were emphasized in Ref.~\cite{CL03_cqg}.
A recent detailed investigation of the conservative 
PN dynamics of spinning compact binaries,
to substantiate the results of Ref.~\cite{JL00},
using again the method of fractal basin boundaries, also observed the
existence of ``chaos'' in these systems \cite{JL03}.
Moreover, it was even shown that 2PN accurate conservative dynamics
of spinning compact binaries can be chaotic even if 
only one of the compact objects spins 
[see Figs.~4 and 5 in Ref.~\cite{JL03}].

In this paper, we show that the conservative 2PN accurate dynamics
of spinning compact binaries \emph{cannot be chaotic},
when only one of the compact objects spins.
We rest our arguments on the fact that 
the above dynamics is integrable.
In fact, it can be shown that even
the conservative 3PN accurate dynamics of
spinning compact binaries,
when spin effects are restricted to the leading order
spin-orbit interaction, is 
integrable in two distinct cases: 
(i) the binary consists of
compact objects of arbitrary mass, where only one of them is spinning with an
arbitrary spin and 
(ii) the binary consists of equal mass
compact objects, having two arbitrary spins.
The integrability is evident from the recent
determination of PN accurate Keplerian-type 
parametric solution to the above two cases \cite{MGS,KG05}.
Our 2PN accurate analytic solutions, 
especially for case (i),
imply that the associated dynamics cannot be chaotic.
However, numerical investigations did report chaos
in 2PN accurate dynamics of spinning compact binaries
when only one of the compact objects spins, i.e.,
our case (i) \cite{JL03}.
We present possible arguments for the above contradiction.

In what follows, we briefly describe 
the recently determined PN accurate parametric solutions
and list their salient features. 
We present our arguments 
why the associated dynamics cannot be chaotic.
We also clarify few erroneous statements made in Ref.~\cite{JL03}.
Finally, we suggest a way to analyze PN accurate dynamics 
of spinning compact binaries, semi-analytically, but in a PN accurate manner.
This should be helpful to determine if 
inspiralling spinning compact binaries,
whose dynamics is correctly given by PN approximation to 
general relativity, can exhibit chaotic behavior, for the cases
not considered here, in a realistic manner.


\section{The relevant binary dynamics and its deterministic nature}

Let us first consider the 2PN accurate conservative dynamics 
of spinning compact binaries, when the spin effects are 
restricted to the leading order spin-orbit interaction.
The dynamics is fully specified by a
PN accurate (reduced) Hamiltonian $H$, which may be
symbolically written as
\begin{align}
\label{eq:1}
{H}(\vek{r}, \vek{p}, \vek{S}_{1}, \vek{S}_{2}) 
& = {H}_{\rm N}(\vek{r}, \vek{p})
+ {H}_{\rm 1PN}(\vek{r}, \vek{p})
+ {H}_{\rm 2PN}(\vek{r}, \vek{p})
\nonumber 
\\
&\quad 
+ {H}_{\rm SO}(\vek{r}, \vek{p}, \vek{S}_{1}, \vek{S}_{2})
\,,
\end{align}
where ${H}_{\rm N}$, ${H}_{\rm 1PN}$, and ${H}_{\rm 2PN}$ are, 
respectively, the Newtonian, first, and second 
PN contributions to the conservative dynamics of
compact binaries, when the spin effects are neglected. The leading
order spin-orbit coupling to the binary dynamics is given
by ${ H}_{\rm SO}$. In the above equation,
$\vek{r} = \vek{\cal R}/(G M)$, $r= |\vek{r}|$, 
and $\vek{p} = \vek{\cal P}/\mu$, where
$\vek{\cal R}$ and $\vek{\cal P}$ are the relative separation vector and
its conjugate momentum vector, respectively.
The  familiar symbols $M$ and $\mu$ have the usual meaning, 
namely, the total mass 
and the reduced mass.
The explicit expressions for 
PN corrections, associated with the motion of nonspinning compact
binaries, were obtained in Refs.~\cite{S85,DS88}.
The spin-orbit contributions, available in Refs.~\cite{KG05,DS88,TD01}, 
read
\begin{align}
{H}_{\rm SO}(\vek{r},\vek{p},\vek{S}_{1},\vek{S}_{2})
= \frac{1}{c^2 r^3} (\vek{r} \times \vek{p}) 
\cdot \vek{S}_\text{eff} 
\,,
\end{align}
where $\vek{S}_\text{eff}$ gives the effective spin
defined by
\begin{align}
\vek{S}_\text{eff} = \delta_{1} \vek{S}_{1} + \delta_{2} \vek{S}_{2}
\,.
\end{align}
In the above  equation,
$\delta_{1} 
= \frac{\eta}{2} + \frac{3}{4} \left(1 - \sqrt{1 - 4 \eta} \right)
$ and 
$\delta_{2} 
= \frac{\eta}{2} + \frac{3}{4} \left(1 + \sqrt{1 - 4 \eta} \right)
$,
where $\eta$ is the finite mass ratio $\eta = \mu /M$.
The reduced spin vectors $\vek{S}_{1}$ and $\vek{S}_{2}$ are related
to the individual spins $\vek{\cal S}_1$ and $\vek{\cal S}_2$ by
$\vek{S}_{1} = \vek{\cal S}_1 /(\mu G M)$ and
$\vek{S}_{2} = \vek{\cal S}_2 /(\mu G M)$, respectively.
We recall that $\vek{\cal R}$, $\vek{\cal P}$, $\vek{\cal S}_1$, and
$\vek{\cal S}_2$ are canonical variables, such that the orbital
variables commute with the spin variables, e.g.,
see Refs.~\cite{Schaefer_grav_effects_2004,TD01}.

Recently, as mentioned earlier, the 3PN accurate Keplerian-type 
parametric solution to the dynamics of spinning compact binaries,
when spin effects are restricted to the leading order 
spin-orbit interaction,
for the following two distinct cases, namely,
(i) the binary consists of compact objects of arbitrary mass, 
where only one of them is spinning with an arbitrary spin and 
(ii) the binary consists of equal mass compact objects, 
having two arbitrary spins,
were obtained in Refs.~\cite{KG05,MGS}.
We display below the 2PN accurate version of the 
above parametric solution.
It turned out that PN accurate dynamics, we are interested
in, allows Keplerian-type parametrization 
in an orbital orthonormal triad,
$(\vek{i},\vek{j},\vek{k})$.
In this triad, $\vek{i}$ defines the line of nodes,
associated with the intersection of the orbital plane with
the invariable plane $(\vek{e}_{X},\vek{e}_{Y})$,
which is the plane perpendicular to 
the total (reduced) angular momentum $\vek{J} = J \vek{e}_{Z}$, 
the only constant vector in our dynamics.
The unit vector $\vek{k}$ is \emph{always}
perpendicular to the orbital plane 
and defined by $\vek{k} = \vek{L}/L$, where 
$\vek{L} = \vek{r} \times \vek{p}$ is the (reduced) orbital 
angular momentum vector. The dynamical vectors of the problem are given by
\begin{align}
\label{eq:sol_for_r_2PN_plus_SO}
\vek{r}(t) & = r (t) \cos\varphi (t) \, \vek{i} (t) + r (t)
\sin\varphi (t) \, \vek{j} (t)
\,,
\\
\vek{L}(t) & = L \vek{k}(t)
\,,
\\
\vek{S}(t) & = J \vek{e}_{Z} - L \vek{k} (t)
\,,
\end{align}
where $\vek{S}$ is the (reduced) total spin, which is related to the 
effective spin by 
$\vek{S}_\text{eff} =  \chi_\text{so} \vek{S} $
for the considered cases.
The mass dependent coupling constant
$\chi_\text{so}$ equals 
$\delta_{1} \; \text{or} \; \delta_{2}$ for the single-spin case
and is given by
$\delta_{1} = \delta_{2} = 7/8 $ for the 
equal-mass case.
The magnitude of $\vek{J}$ is denoted by $J$ and is given by
$J = (L^2 + S^2 + 2 L S \cos\alpha)^{1/2}$. Note that 
the angle $\alpha$ between $\vek{L}$ and $\vek{S}$ can be 
chosen quite
arbitrarily.
The time dependent basic vectors $(\vek{i},\vek{j},\vek{k})$ are
explicitly given by
\begin{subequations}
\label{eq:ijk_in_the_combined_sol}
\begin{align}
\vek{i} & =
\cos\Upsilon  \vek{e}_{X}
+ \sin\Upsilon  \vek{e}_{Y}
\,,
\\
\vek{j} & =
- \cos\Theta \sin\Upsilon  \vek{e}_{X}
+ \cos\Theta \cos\Upsilon  \vek{e}_{Y}
+ \sin\Theta \, \vek{e}_{Z}
\,,
\\
\vek{k} & =
\sin\Theta \sin\Upsilon \vek{e}_{X}
-\sin\Theta \cos\Upsilon \vek{e}_{Y}
+\cos\Theta \, \vek{e}_{Z}
\,.
\end{align}
\end{subequations}
The angle $\Theta$ gives the constant precessional angle between
$\vek{L}$ and $\vek{J}$ [see Fig.~1 in Ref.~\cite{KG05}].
The time evolution for $r$, $\varphi$, and $\Upsilon$ is
given in a parametric and PN accurate way, which reads  
\begin{subequations}
\label{eq:FinalPara_ADM_2PN_plus_SO}
\begin{align}
r & 
= a_r \left( 1 - e_r \cos u \right )
\,,
\\
l \equiv n \left( t - t_0 \right) & 
= u - e_t \sin u 
+ \frac{g_{4t}}{c^4} (v - u)
+ \frac{f_{4t}}{c^4} \sin v
\,,
\\
\varphi - \varphi_{0} & 
= (1 + k) v
+ \frac{f_{4\varphi}}{c^4} \sin 2 v
+ \frac{g_{4\varphi}}{c^4} \sin 3 v
\,,
\\
\label{eq:Ups_min_Ups0_in_the_comb_sol}
\Upsilon - \Upsilon_0 & 
= \frac{\chi_\text{so} J}{c^2 L^3} ( v + e \sin v )
\,,
\\
\text{where} \quad
v & 
= 2\arctan \left[ \left( \frac{ 1 + e_{\varphi}}{ 1 - e_{\varphi}}
\right)^{1/2} \tan \frac{u}{2} \right]
\,.
\end{align}
\end{subequations}
The PN accurate orbital elements $a_r$, $e_r^2$, $n$, $e_t^2$ , $k$,
and $e_{\varphi}^2$, and the PN order orbital functions
$g_{4t}$, $f_{4t}$, $f_{4\varphi}$, and $g_{4\varphi}$,
expressible in terms of $E$, $L$, $S$, $\eta$, and $\alpha$, 
are obtainable from Ref.~\cite{KG05}.
We recall that the above parametric solution is usually referred to as
the generalized quasi-Keplerian parametrization for
the PN accurate orbital dynamics of compact binaries
and extends the seminal works done by Damour, Deruelle, Sch\"afer,
and Wex \cite{DD85,DS88,SW93,W95}.
The quasi-Keplerian parametric solution is heavily employed to
construct the ``timing formula'' for relativistic binary pulsars,
which in turn is used to extract astrophysical information from these systems
and to test general relativity in strong field regimes \cite{DD86,DT91}.
The recent determination of ``ready to use'' search templates for 
nonspinning compact binaries moving in inspiralling 
eccentric orbits, given in Ref.~\cite{DGI},
also required the above mentioned 
quasi-Keplerian parametrization.

 The existence of such a PN accurate parametric solution and
hence the integrability of the spinning binary dynamics
should not be surprising if one \emph{carefully} counts the 
degrees of freedom and the existing conserved quantities 
in the above mentioned two distinct cases,
where our parametrization is valid.
In our Hamiltonian system, for the two cases considered, the
dimension of the associated phase space is $8$
[the nonspinning part of the Hamiltonian contributes $6$ dimensions
and the spin contributes $2$ dimensions.
This indicates that the number of independent degrees of freedom,
for the two cases considered,
is $4$ and there are $4$ independent constants of motion,
namely, $E$, $L$, $S$, and $\vek{L} \cdot \vek{S}$.
The deterministic nature of the Hamiltonian, given by Eq.~\eqref{eq:1} 
and treated as a self-contained dynamical system, naturally follows
from above arguments.

One may argue, that the equations of motion for spinning compact binaries
presented in Ref.~\cite{K95} and employed in Refs.~\cite{JL03,JL00}
are in harmonic coordinates and employ 
a different spin supplementary condition (SSC)
than the one used in Ref.~\cite{KG05}.
Therefore it is not desirable to compare results of Ref.~\cite{JL03} 
with results of Ref.~\cite{KG05}, especially about the 
deterministic nature of the underlying dynamics.
Following Ref.~\cite{MGS}, it is trivial to obtain a 
harmonic coordinate version of 
Eqs.~\eqref{eq:sol_for_r_2PN_plus_SO}--\eqref{eq:FinalPara_ADM_2PN_plus_SO}. 
Furter, since different SSCs,
which may be regarded as different ways to specify center-of-motion
world line of each spinning body, can be related by a PN order
shift of the center-of-motion world lines 
[see Appendix~A of Ref.~\cite{K95}],
it is clear that the equations of motion, given in Ref.~\cite{K95} 
and that given by Eq.~\eqref{eq:1}, for the two cases considered, 
describe the same dynamics.
More importantly, 
using Eqs.~(2.6)--(2.13) of Ref.~\cite{K95},
it is straightforward to show that
the quantities $E$, $L$, $S$, and $\vek{L} \cdot \vek{S}$ 
are also conserved to the required PN order
for our two cases.
This shows that the deterministic nature
of the dynamics of spinning compact binaries,
for the two distinct cases (i) and (ii),
does not depend on coordinate conditions, SSCs, 
and the representations.
It is interesting to note (again)
that Ref.~\cite{JL03} reported chaos in case (i),
where only one object spins,
using the dynamics prescribed in Ref.~\cite{K95}.

The physical argument employed in Ref.~\cite{JL03} to explain
the apparent ``chaos'' is the highly irregular behavior of the orbital plane
due to the inclusion of leading order
spin effects on to the conservative
2PN accurate dynamics, associated with
the nonspinning compact binaries.
In Ref.~\cite{JL03}, the Newtonian orbital angular momentum vector
was used to specify the orientation of the orbital
plane. We believe that it is not an appropriate quantity to specify 
the orbital plane as it is \emph{not even} perpendicular
to the orbital plane, when spin effects are included.
However, in the Hamiltonian formulation of PN dynamics,
$\vek{L} \equiv \vek{r} \times \vek{p}$, always remains perpendicular
to the orbital plane and easily tracks its orientation.
Recall that $\vek{L}$, as defined above,
also occurs in the Poincar\'e algebra of
gravitating two-body system of spin-less 
particles to 3PN order \cite{DJS00a}.
We observe that in the two cases, where analytic solutions 
exist, the orbital plane, specified by $\vek{L} $,
carves out a simple cone as it precesses around the invariant
direction, defined by the total angular momentum $\vek{J}$.
Further note that, in these cases,
$\dot{\vek{L}} = - \dot{\vek{S}}$ and we think this is not the
way the evolution of $\vek{L}$ and $\vek{S}$ is treated in
Ref.~\cite{JL03} [see Eq.~(2.6) in Ref.~\cite{JL03}].
When spin-spin effects are included,
the angle between $\vek{L}$ and $\vek{J}$ no longer 
remains a constant, as is roughly evident 
from Eq.~(4.15) in Ref.~\cite{KG05}.
However, we note that the time scale for the evolution of
$\Theta$, which defines the angle between $\vek{L}$ and $\vek{J}$,
should be large and the amplitude of its variation small.
This is so as spin-spin interactions, responsible for
the variation of $\Theta$,
may be treated to be 2PN or 3PN order corrections,
depending on the magnitudes of spins
[see discussions at the end of Sec.~II in Ref.~\cite{KG05}].

It was stated in Ref.~\cite{JL03} that beyond 2PN order, the dynamics
of compact binaries will be chaotic even in the absence of spin.
The erroneousness of this conclusion is clearly
evident from the recent determination of 3PN accurate 
generalized quasi-Keplerian parametrization for
the solution of the 3PN accurate
equations of motion for two nonspinning
compact objects moving in eccentric orbits \cite{MGS}.
The underlying dynamics,  derivable from an 
ordinary Hamiltonian, can be extracted from the papers
of Damour, Jaranowski, and Sch\"afer \cite{DJS01}.
It can easily be shown that the motion is restricted to a plane, namely the
orbital plane, and we can introduce polar coordinates 
such that $\vek{r} = r ( \cos \varphi, \sin \varphi, 0)$.
The 3PN accurate generalized quasi-Keplerian
parametrization for compact binaries moving in eccentric orbits, 
in the center-of-mass frame, is then given by
\begin{widetext}
\begin{subequations}
\label{e:FinalParamADM}
\begin{align}
r & = a_r \left( 1 - e_r \cos u \right)
\,,
\\
l \equiv n \left( t - t_0 \right) & 
= u - e_t \sin u 
+ \bigg( \frac{g_{4t}}{c^4} + \frac{g_{6t}}{c^6} \bigg) (v - u)
+ \bigg( \frac{f_{4t}}{c^4} + \frac{f_{6t}}{c^6} \bigg) \sin v
+ \frac{i_{6t}}{c^6} \sin 2v
+ \frac{h_{6t}}{c^6} \sin 3v 
\,,
\\
\varphi - \varphi_{0} & 
= (1 + k ) v 
+ \bigg( \frac{f_{4\varphi}}{c^4} + \frac{f_{6\varphi}}{c^6} \bigg) \sin 2v
+ \bigg( \frac{g_{4\varphi}}{c^4} + \frac{g_{6\varphi}}{c^6} \bigg) \sin 3v
+ \frac{ i_{6\varphi} }{c^6} \sin 4v
+ \frac{ h_{6\varphi} }{c^6} \sin 5v 
\,,
\\
\text{where} \quad
v & = 2 \arctan 
\left[ 
\left( \frac{ 1 + e_{\varphi} }{ 1 - e_{\varphi} } \right)^{1/2} 
\tan \frac{u}{2} 
\right]
\,.
\end{align}
\end{subequations}
\end{widetext}
The PN accurate orbital elements $a_r$, $e_r^2$, $n$, $e_t^2$ , $k$,
and $e_{\varphi}^2$, and the PN orbital functions
$g_{4t}$, $g_{6t}$, $f_{4t}$, $f_{6t}$, $i_{6t}$, $h_{6t}$,
$f_{4\varphi}$, $f_{6\varphi}$, $g_{4\varphi}$, 
$g_{6\varphi}$, $i_{6\varphi}$, and $h_{6\varphi}$, 
expressible in terms of $E$, $L$, and $\eta$,
are obtainable from Ref.~\cite{MGS}.
The number of degrees of freedom are three
and there are sufficient conserved quantities to restrict
the orbits to a plane and the dynamics to be integrable.
The deterministic nature of the dynamics 
does not change when radiation reaction,
which enters at 2.5PN order, is included.
It is actually possible to incorporate, almost analytically,
the effects of reactive ${H}_{\rm 2.5PN}(\vek{r},\vek{p})$,
derivable easily from Ref.~\cite{S85}, in to the 3PN accurate conservative 
dynamics of Ref.~\cite{MGS}, employing
the techniques described in Ref.~\cite{DGI}.
We would like to remind that a fully 2.5PN accurate orbital
dynamics of inspiralling binaries consisting of nonspinning compact objects
of arbitrary mass ratio, available in Ref.~\cite{DGI}, requires
numerical evaluation of just one quadrature
[see Sec.~V of Ref.~\cite{DGI} and discussions therein].
Further, with the help of Refs.~\cite{MGS,DGI},
one can easily conclude that the motion is still restricted to a plane
to 3PN order in the case of inspiralling nonspinning 
compact binaries.

Finally, note that PN accurate differential equations,
employed in Ref.~\cite{JL03}, are \emph{realistic}, 
only if one is allowed to neglect the omitted
higher order correction terms.
Therefore, extreme care should be taken to make sure that 
the solutions to these differential equations are also as PN 
accurate and this is not an easy task,
while solving these differential equations numerically.
In the parametric solutions, described above, we can easily
control their PN accuracy and determine the validity 
of the approximate dynamcis in a gauge invariant way. 
However, it is not always possible to obtain 
PN accurate simple parametric solutions, 
as given in Refs.~\cite{MGS,KG05}.
For example, in case of a spinning compact binary
with leading order spin-orbit interaction,
it will not be possible to obtain a 
simple parametric solution for the general case,
where $m_{1} \neq m_{2}$ and $\vek{S}_1 \neq \vek{S}_2$,
as $\Theta$ is no longer a constant angle there.
In this case, it is still possible to parametrize the radial motion,
but the angular motions do not accommodate any simple parametric solutions.

We want to point out that
our case (ii), which includes two spinning objects
with leading order spin-orbit interaction,
should not be taken to argue that
spin-spin interactions will not make the dynamics
of spinning compact binaries chaotic.
Our case (ii) should be viewed as a toy
model to test numerical investigations that
probe the deterministic nature of PN accurate dynamics
of spinning compact binaries.
When ${H}_{\rm SS}$ terms, which specify the spin-spin interactions,
are included in Eq.~\eqref{eq:1},
even the radial motion does not allow simple parametrization.
However, it should be possible to obtain semi-analytic 
PN accurate solution to the PN accurate dynamics of spinning compact
binaries, moving in eccentric orbits, when all leading order spin
effects are included.
This will require adapting techniques from classical perturbation theory
and will be similar to the way 
classical spin-orbit coupling was incorporated into the Newtonian
accurate orbital motion \cite{R78,W98}.
We believe that such PN accurate solutions to PN accurate binary
dynamics may be more suitable to analyze the question of 
``chaos'' in PN accurate dynamics of spinning compact binaries,
in a realistic manner. Many of these issues are currently under
investigation in Jena.


\section{Summary}

Let us recapitulate. We have shown that the conservative
PN accurate dynamics of spinning compact binaries
with leading order spin-orbit interaction, 
for the two distinct cases mentioned above,
is integrable and hence cannot be chaotic.
We emphasized that these two scenarios are as deterministic
as the PN accurate conservative dynamics of nonspinning compact binaries.
We hope that our parametric solutions,
for the two distinct cases, will be employed to test
numerical investigations that probe the question
of chaos in spinning compact binaries.
Finally, we feel that PN accurate semi-analytic solutions 
to the PN accurate dynamics of 
spinning compact binaries may help to clarify
the question of chaos in spinning compact binaries.


\begin{acknowledgments}

We are grateful to Gerhard Sch\"afer for 
illuminating discussions and persistent encouragement.
Conversations with T.~Damour and R.-M.~Memmesheimer 
are warmly acknowledged.
The work is supported by the Deutsche Forschungsgemeinschaft (DFG)
through SFB/TR7 ``Gravitationswellenastronomie''.

\end{acknowledgments}


\end{document}